%% file: template.tex
\documentclass{article}

\usepackage{arxiv}

\usepackage[utf8]{inputenc} % allow utf-8 input
\usepackage[T1]{fontenc}    % use 8-bit T1 fonts
\usepackage{hyperref}       % hyperlinks
\usepackage{url}            % simple URL typesetting
\usepackage{booktabs}       % professional-quality tables
\usepackage{amsfonts}       % blackboard math symbols
\usepackage{nicefrac}       % compact symbols for 1/2, etc.
\usepackage{microtype}      % microtypography
\usepackage{lipsum}

\usepackage{amsmath,amssymb}
\usepackage[squaren]{SIunits}
\usepackage{multirow}
\usepackage{adjustbox}
\usepackage{graphbox}

\title{Characterization of \emph{Posidonia Oceanica}  Seagrass Aerenchyma through Whole Slide Imaging: A Pilot Study}

\author{
  Olivier Debeir \\
  Laboratory of Image Synthesis and Analysis (LISA)\\
  Université Libre de Bruxelles (ULB)\\
  Brussels, Belgium \\
  \texttt{odebeir@ulb.ac.be} \\
  \And
  Justine Allard\\
  DIAPath, Center for Microscopy and Molecular Imaging (CMMI)\\
  Gosselies, Belgium\\
  \And
  Christine Decaestecker \\
 Laboratory of Image Synthesis and Analysis (LISA)\\
  Université Libre de Bruxelles (ULB)\\
  Brussels, Belgium \\
  \And
  Jean-Pierre Hermand\thanks{\dag In memoriam to J.-P. Hermand who passed away suddenly on September 3, 2018.} \\
  Environmental HydroAcoustics Lab (EHL)\\
  Université Libre de Bruxelles (ULB)\\
  Brussels, Belgium \\
}

\begin{document}
\maketitle

\begin{abstract}
Characterizing the tissue morphology and anatomy of seagrasses is essential to predicting their acoustic behavior. In this pilot study, we use histology techniques and whole slide imaging (WSI) to describe the composition and topology of the aerenchyma of an entire leaf blade in an automatic way combining the advantages of X-ray microtomography and optical microscopy. Paraffin blocks are prepared in such a way that microtome slices contain an arbitrarily large number of cross sections distributed along the full length of a blade.
The sample organization in the paraffin block coupled with whole slide image analysis allows high throughput data extraction and an exhaustive characterization along the whole blade length.
The core of the work are image processing algorithms that can identify cells and air lacunae (or void) from fiber strand, epidermis, mesophyll and vascular system.
A set of specific features is developed to adequately describe the convexity of cells and voids where standard descriptors fail. The features scrutinize the local curvature of the object borders to allow an accurate discrimination between void and cell through machine learning.
The algorithm allows to reconstruct the cells and cell membrane features that are relevant to tissue density, compressibility and rigidity. 
Size distribution of the different cell types and gas spaces, total biomass and total void volume fraction are then extracted from the high resolution slices to provide a complete characterization of the tissue along the leave from its base to the apex.
\end{abstract}

% keywords can be removed

\section{Introduction}
Seagrasses play a major role in coastal ecosystems of all continents except Antarctica, due to their high biological productivity and environmental conservation \cite{mtwana16}.
They form densely‑populated communities on various subtidal and intertidal substrates, hosting a wide variety of inhabitants.
Worldwide, they occupy \unit{600,000}{\kilo\squaremetre} of the continental shelf, contributing $12\%$ of the total carbon storage in the ocean.
Despite three decades of intensive research on biology and ecology of seagrasses, very few studies have been undertaken to study their acoustic behavior, much remaining unknown about the species‑specific, acoustic properties of their tissue \cite{johnson17c}.
Such knowledge is nevertheless essential for mapping their habitats with echosounders and for studying primary production through sound propagation measurement at the meadow scale \cite{hermand03}.
Early field experimentation \cite{hermand98} and laboratory investigation \cite{wilson10} have shown that tissue porosity is a key determinant of the acoustics of seagrasses but remains insufficient to model their behaviour of a wide range of frequencies. In recent years, evidence has been obtained that cell composition and structure of the leaf blade aerenchyma determine the acoustic properties through investigation of various species subjected to acoustic resonance \cite{johnson17b} and ultrasonic measurements\cite{johnson17a}.
A more complete model with detailed information on the seagrass morphology and anatomy is needed which accounts for shape, cell composition, and tissue characteristics.

In spite of their typically thin tissue, seagrasses have a morphology liable to produce an acoustic signature (e.g., gas channels running along the leaf blade), which motivates the need for detailed models that account for all acoustically-relevant tissue features. Traditionally, the void fraction profile along a leaf blade can be measured by macroscopic means such as buoyancy. However, such measurements do not provide information on the morphology of the gas channels running along the blade, although their equivalent diameter and spacing is important to acoustics.
Beyond such a coarse - though important \cite{johnson18a} - description of a plant tissue, modern imaging types, including laser scanning confocal microscopy \cite{hasselhoff03} and X-ray synchrotron imaging, \cite{mebatsion09} enable a 2D or 3D reconstruction with very high resolution. However, their very limited field of view requires a lot of image acquisitions to achieve a wide specimen screening. For example, synchrotron resolution ranges from \unit{0.3}{\micro\meter}--\unit{5}{\micro\meter} for fields of view from \unit{1}{\milli\meter}--\unit{20}{\milli\meter}, a resolution of \unit{5}{\micro\meter} being insufficient for the problem at hand.
Depending on species, other imaging modalities used in biomedical applications do not have sufficient resolution.
Recent advances in X-ray microtomography provide more and more detailed volumetric images. MicroCT works particularly well on porous soft tissue such as aerenchyma whose overall low density results in low absorption and a high density contrast between cell and void enables clear reconstruction of the gas channels which determine acoustic behavior. Micro-CT resolution range typically from \unit{5}{\micro\meter}--\unit{50}{\micro\meter} corresponds to field of view from \unit{1}{\centi\meter}--\unit{20}{\centi\meter}.
However, the low density contrast between cell types does not allow to resolve other mechanically important feature such as fiber strands.
Whole slide imaging (WSI) provides a large number of high-quality cross-sectional images with a high resolution (20X or 40X) which cover the whole leaf blade with a single block of paraffin or resin.

In this paper, images acquired by WSI are processed by algorithms developed to study in a quantitative and exhaustive way the tissue arrangement and cell shape. Both structures have an effect on acoustics, the anatomy of the inter-cellular gas-filled spaces of aerenchyma being of particular interest. The studied species is an endemic phanerogam of the Mediterranean Sea. It has long ribbon shaped leaves, grouped in shoots, which develop on various substrates in \unit{1}{\metre} to \unit{50}{\metre} water depths, depending mainly on light availability. For the purpose of this work, we focus on the leaf-blade aerenchyma of \emph{Posidonia oceanica}, one of the three main seagrass species of the Mediterranean Sea. Besides of its high ecological importance, the characteristic small size of air lacunae and marked variability in microstructure and composition along the blade make the analysis challenging.

\section{Materials and Methods}
\subsection{Plant species and origin}

\begin{figure}[ht]
\centering
\begin{tabular}{cc}
   A. \includegraphics[height=2.0in]{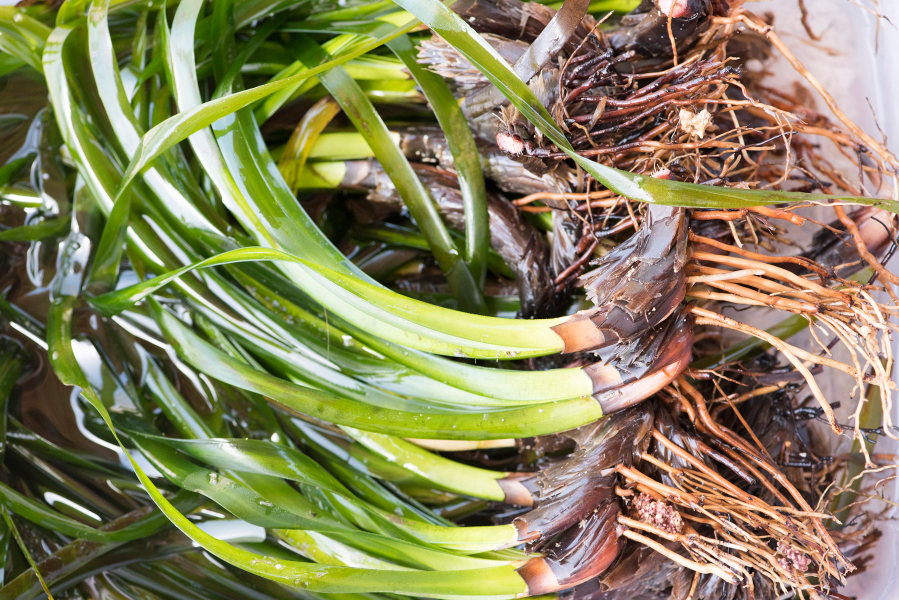} & B. \includegraphics[height=2.0in]{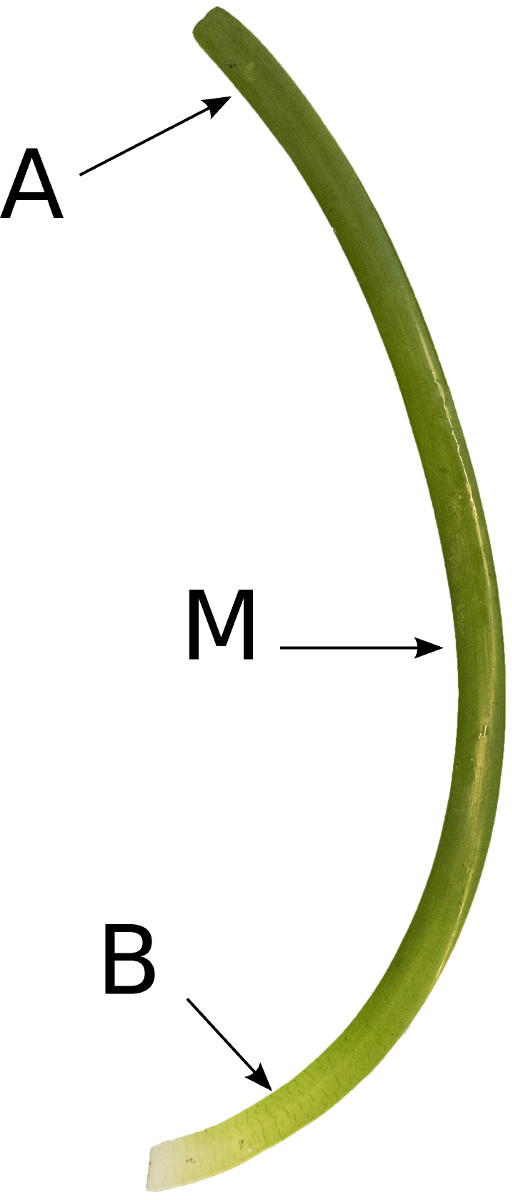}\\
   \multicolumn{2}{c}{C. \includegraphics[height=3.0in]{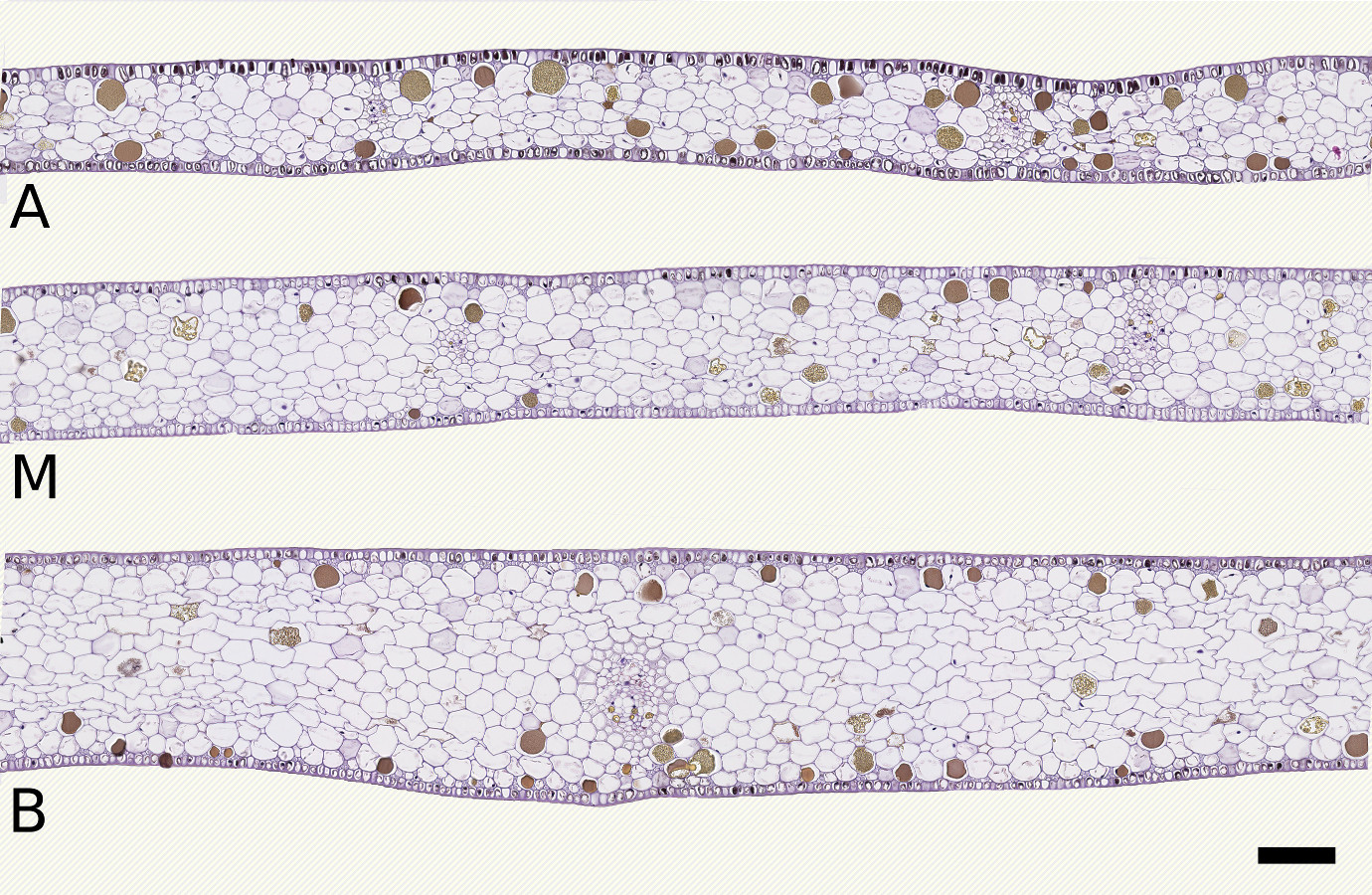}}
   \\
\end{tabular}
\caption{\textbf{Sample description.}
A: Freshly collected \emph{P. oceanica} seagrasses including above- and below-the-ground tissue. B: Sampling sites of the investigated leaf blade: base (B), middle (M) and apex (A). C: Brightfield microscopic view of one typical cross-section at each site as extracted from a whole slide image (scale bar = 100{\micro\meter}). Magnification: 20X (pixel size \unit{0.46}{\micro\meter}x\unit{0.46}{\micro\meter}).}
\label{fig_material}
\end{figure}

The investigated samples of \textit{Posidonia~oceanica} seagrass species originate from coastal waters in North Western Crete, Greece. Whole plants in very healthy conditions were collected on a rocky substrate in \unit{5}{\metre} water depth amounting to 10+ shoots of 6–-7 stems (\figurename~\ref{fig_material}A).
The plants were stored with a small volume of seawater in an ice-refrigerated container and transported to our laboratory in Brussels.

\subsection{Sample fixation and cross-sectioning}

A leaf blade almost free of epiphyte was selected and fixed 36 hours in $4\%$-buffered formaldehyde. Three sampling sites along the blade with most distinct characteristics were deemed sufficient to develop our algorithms.
The chosen sites are the white-greenish tissue at the base just above the sheath, the greener tissue in the middle and the darker green tissue near the apex, a color trend reflecting an increase in chloroplast concentration toward the apex (\figurename~\ref{fig_material}B). The microscopic view of the corresponding cross‑sections (\figurename~\ref{fig_material}C) shows morphological variability along the leaf blade, from thick mature tissue near the base to thin younger tissue toward the apex.
To describe tissue microstructure and composition - and their variations along a leaf blade - the number of cross-sections at each site was chosen to achieve a good trade-off between statistical confidence and axial resolution.
At each site, the blade was sectioned transversely into a number of contiguous segments of \unit{5}{\milli\meter} in length which were then stacked with space interleave to fit the embedding block. Such arrangement enables fast simultaneous processing of a larger tissue sample originating from different locations of the same blade.

\begin{figure}[ht]
\centering
\includegraphics[width=5.0in]{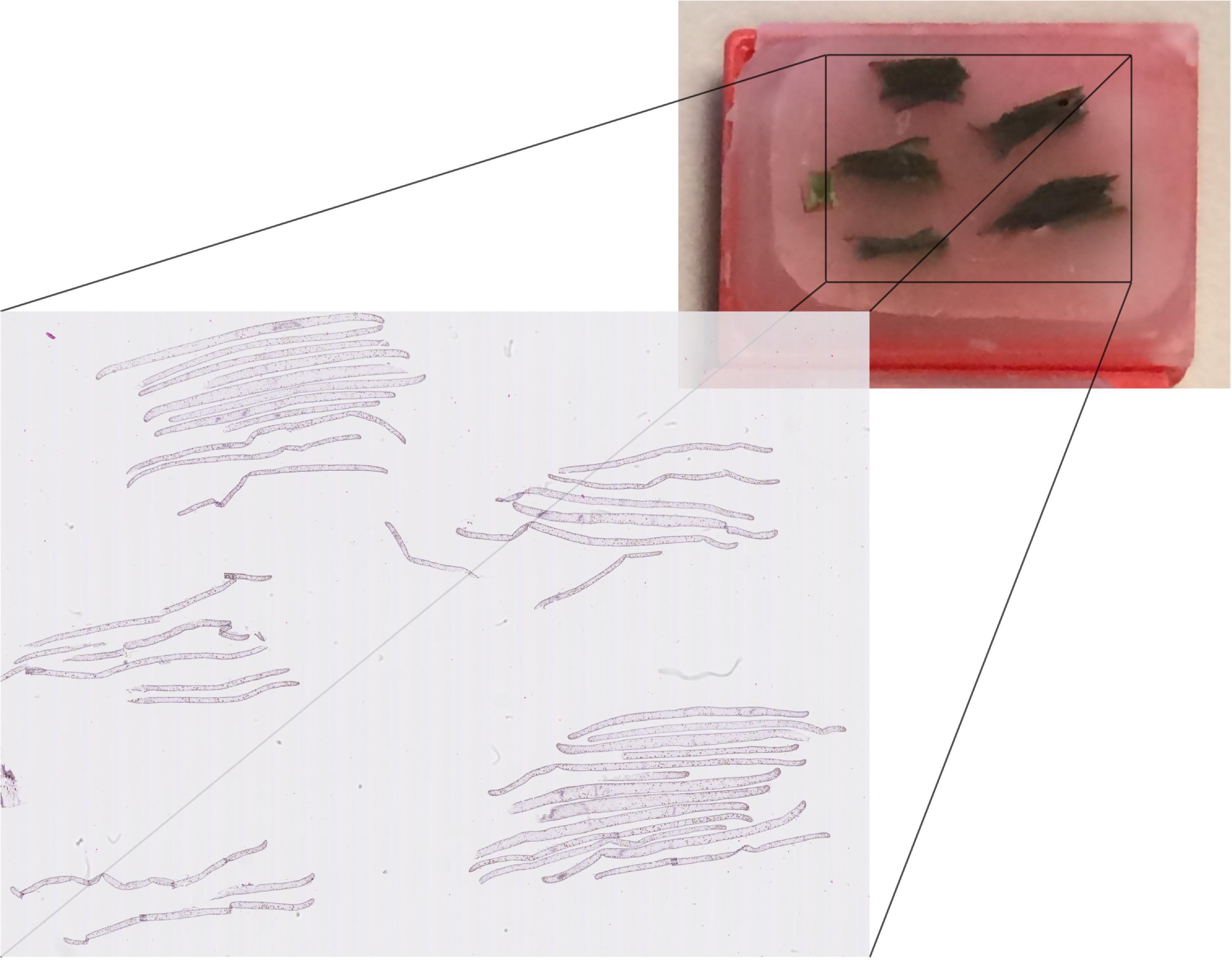}
\caption{\textbf{Blade segment stacking.}
Leaf blade segments are arranged into small stacks to increase sampling on a single slide. For this block, the segments were not regrouped according to their position along the blade. Nevertheless, the respective positions were easily determined from basic anatomical features such as the number of cell layers.}
\label{fig_lowres} 
\end{figure}

\subsection{Paraffin embedding}

Paraffin embedding was tested as an alternative to resin because of faster handling and reduced cost for volume processing. Although its occasional difficulties on delicate tissue, the paraffin method is far less tedious and expensive than the resin method. Very fresh specimens of the strong tissue of \emph{P. oceanica} tissue appear to resist to both this method and the following treatment.

\subsection{Block slicing and staining}

A \unit{5}{\micro\meter}-thick slice was cut from the paraffin block (\figurename~\ref{fig_lowres}), placed on a microscope glass slide and then baked at a temperature of 37\celsius overnight. The tissue slice was then stained by a combination of hematoxylin and eosin (HE) solutions using a Tissue-Tek DRSTM autostainer (Sakura Finetek, Netherlands) and the following standardized protocol. A xylene solution (2\,x\,4\,min) removes the paraffin from the tissue which is then re-hydrated with graded alcoholic solutions (3\,x\,1.15\,min). The HE staining colors in blue the cell nuclei, by a treatment with sodium tetraborate (3\,min), and counterstains the tissue in pink with an alcoholic solution of eosin (20 sec). As compared to toluidine, which is commonly used for plants, HE was found to provide the right level of detail to delineate the exact position of the cell boundaries. Finally, the slide is rinsed with demineralized water, dehydrated using alcohol, mounted with Entellan and coverslipped.

\subsection{Whole slide scanning}

The slide was scanned at 20X magnification (NanoZoomer HT 2.0, Hamamatsu, Hamamatsu City, Japan). The resulting image size is 53760\,x\,39936 pixels with a pixel size of \unit{454}{\nano\metre} x \unit{454}{\nano\metre}. The image is stored as three 8-bit channels (RGB), $80\%$ compressed JPEG in a TIFF-like proprietary format (NDPI). \figurename~\ref{fig_leaf_morphology} shows examples of sections at base, mid and apex positions along the blade (\figurename~\ref{fig_material}B), as extracted from the whole slide image. HE staining highlights the cell boundaries (membrane) but does not provide unnecessary details of the cell content besides its stained blue nucleus when present in the section. The phenolic compounds are naturally brown colored.

\subsection{Tissue structure and objects of interest}

\begin{figure}[ht]
\centering
\includegraphics[width=5.0in]{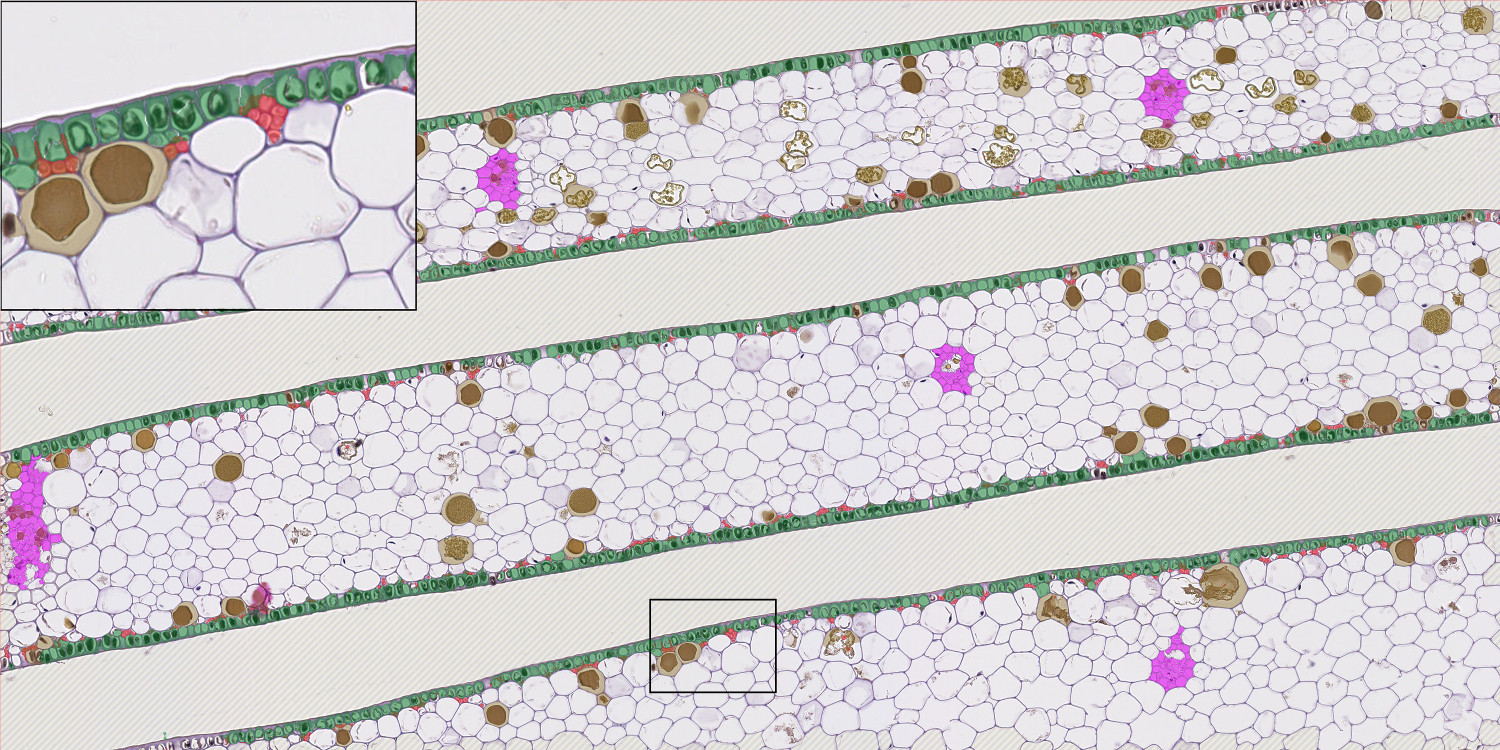}
\caption{\textbf{Bright microscopy image of \emph{P. oceanica} leaf blade.}
Transverse sections are shown at the positions A, B and C indicated in \figurename~\ref{fig_material}A. The hatched area is the slide background. The green overlay highlights the epidermis. Strands of fiber cells are evidenced in red color. The mesophyll tissue is not colored except for the vascular bundles overlaid with purple and 'phenolic' cells overlaid in brown, respectively.}
\label{fig_leaf_morphology} 
\end{figure}

The length of the investigated leaf blade is \unit{0.6}{\metre} and its width and thickness range from \unit{\approx 4.5-9}{\milli\metre} and \unit{\approx 150-340}{\micro\metre}, respectively (\figurename~\ref{fig_leaf_morphology}). The photosynthetic apparatus of \emph{P. oceanica} allows for maximum release of photosynthetic oxygen to the ambient medium \cite{colombo83}. The leaf blade consists of a monolayered epidermis and a mesophyll whose number of layers varies from base to apex (from four to 12 layers in the studied sample). The \unit{0.5}{\micro\meter}-thick cuticle is not visible at this 20X magnification. In the epidermis, which is the major site of photosynthesis, chloroplasts are densely arranged in small, radially elongated cells. A strand of fiber cells - barely visible at this magnification - are evidenced by a red overlay. The lacunar system is constituted by interconnected air channels and small pores within the mesophyll. Its particularly small dimensions are a distinctive feature of \emph{P. oceanica}.  The air lacunae which are called 'void' in the rest of the paper are not so easily distinguished from the cells. In contrast, the veins and phenol storing cells are easily recognized.
The following objects should be automatically extracted: the fiber cells, the voids, the cells, the veins, the epithelium, the phenolic compound.

\subsection{Low-resolution / full frame image processing}

Regarding the high native image resolution and size, image processing is to be done at two different resolutions. The low resolution allows to process the full slice in a single field of view but with a very limited resolution (as illustrated in \figurename~\ref{fig_lowres}) that corresponds to a 1/16
sub-sampling for which the whole slide fits in a 3360 x 2496-px image. 

The low resolution image processing aims to separate the tissue pixels from the background. Tissue is identified from the background by thresholding the optical density (OD) of the grayscale image (using an arbitrary value of .1) and constitutes the Tissue mask. A Label image is built by labelling each connected component of tissue as one blade piece, this enables to group further segmented structures by blade piece, i.e. by its position along the original leave (i.e. base, middle and apex). A morphological gradient is applied to the Tissue mask to identify the epithelium of each blade, the radius of the gradient filter is chosen accordingly to the observed average epithelium thickness (5 pixels in the low-resolution image); the resulting image is the Epithelium mask.

\subsection{Full-resolution / tiles processing}

For the processing at full resolution, the scan image was partitioned into 4096 x 2048-px tiles to limit memory requirement. The full resolution tile processing aims to identify each object composing the blade tissue. 

\subsubsection{Phenolic pixel removal}

Basically, all the objects forming the tissue are clusters of bright pixels separated by denser pixels of cell membranes. Some cells exhibit an opaque phenolyc content that makes it more complex to distinguish between objects and membranes. Therefore we identify and replace phenolic pixels by background-like pixels. A low-level processing is applied to split the RGB color components present in the HE‑stained tissue by means of `color deconvolution’ to extract three channels corresponding to hematoxilyn (blue), eosin (pink) and DAB (brown), as proposed by Ruifrok et al. \cite{ruifrok01}.
The brown/DAB channel allows to simply identify phenolic pixels with an OD threshold (arbitrarily set to .1), giving the Phenolic mask. The OD of the blue channel is used as the Tissue density image where high values correspond to membranes and low value to background. The pixels of the Tissue density image being positive in the Phenolic mask are replaced by 0 (i.e. null OD). It is this corrected version of the Tissue density image which is considered for further processing described below.

\subsubsection{Object segmentation}

\begin{figure}[ht]
\centering
\begin{tabular}{cc}
   A. \includegraphics[height=2.0in]{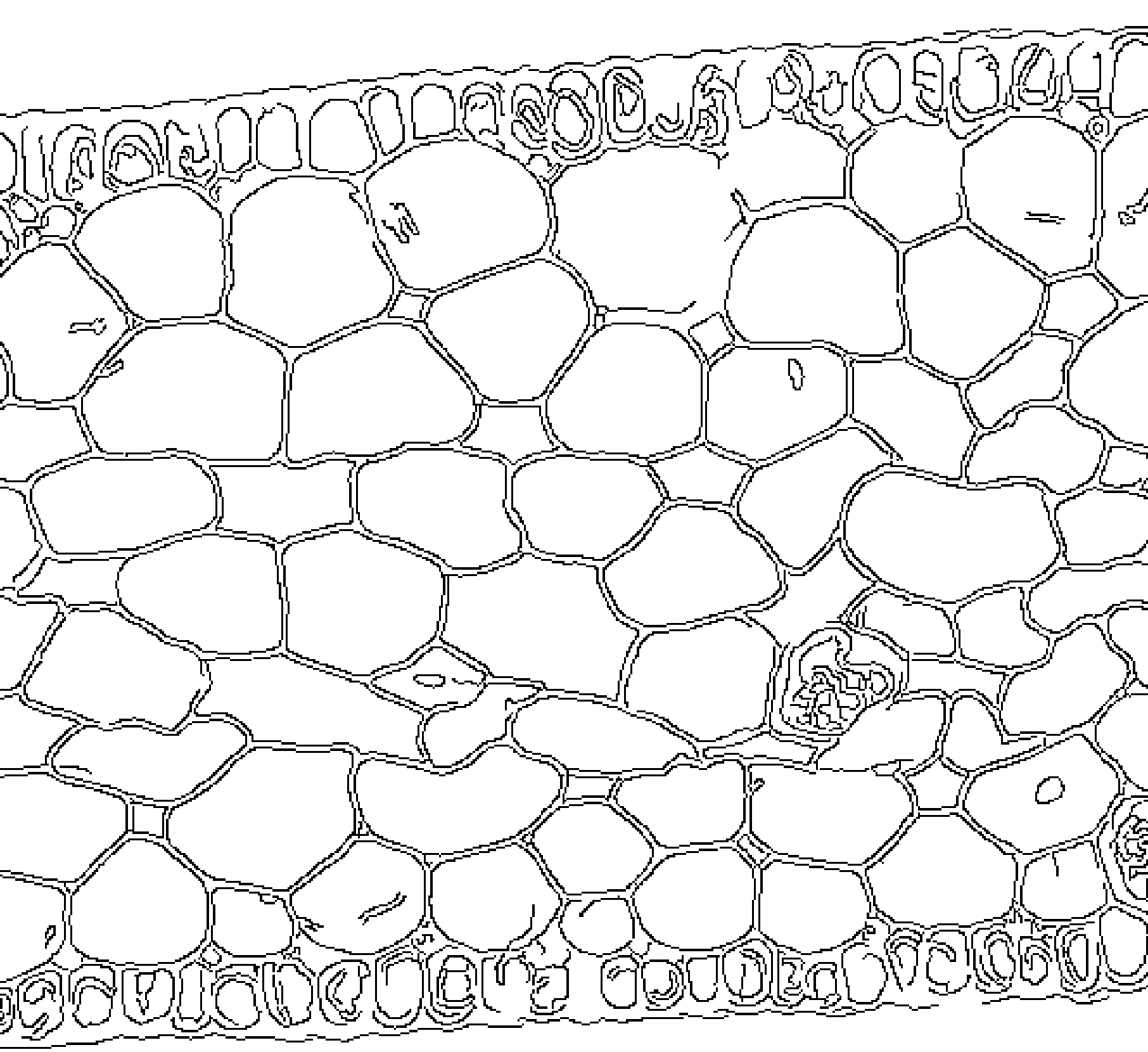} & B. \includegraphics[height=2.0in]{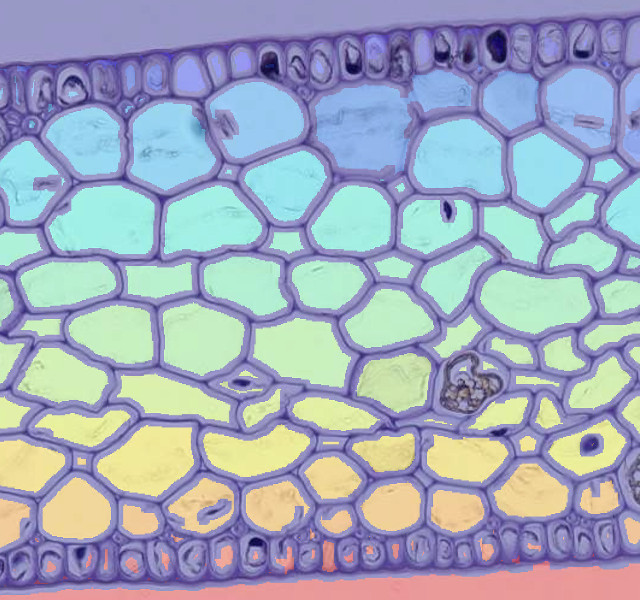}\\
   C. \includegraphics[height=2.0in]{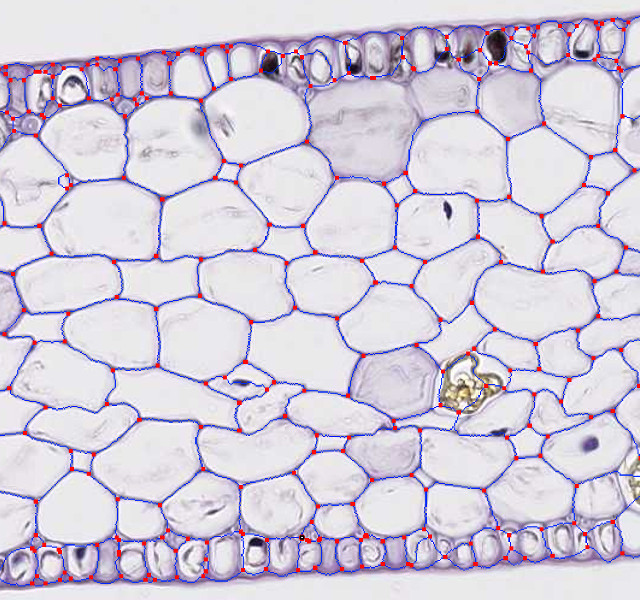} & D. \includegraphics[height=2.0in]{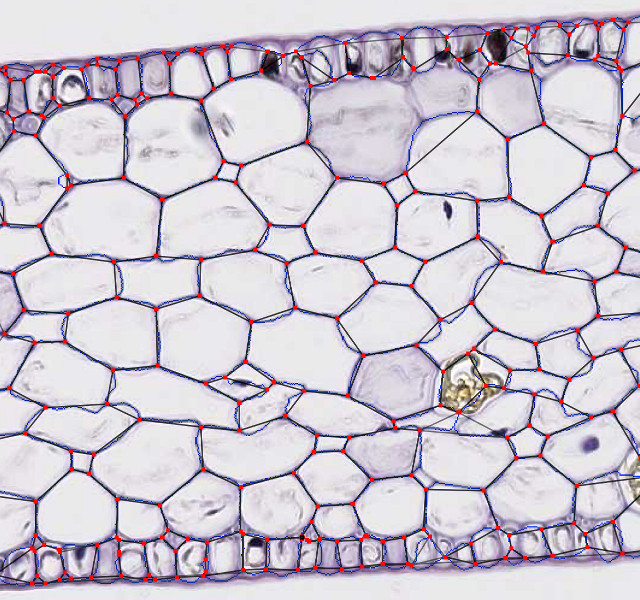}\\
  
\end{tabular}
\caption{{\bf Aerenchyma segmentation.}
A: Canny edge, B: object marker, C: membrane triple points and D: straight line segment joining triple points. The objects can be cells or voids.}
\label{fig_seg}
\end{figure}

Canny’s edge detection \cite{canny86} on the Tissue density image delineates all object borders (\figurename~\ref{fig_seg}A). Similarly to \cite{peachey90} and \cite{travis96}, an Euclidian distance map \cite{borgefors86} is computed based on the detected Canny’s borders. In the resulting map the further away a pixel is from the Canny’s edges, the higher its value is. The Markers image is then build by thresholding the distance map above a value of three pixels, that corresponds to the observed half thickness of the cell membranes. This threshold ensures that markers are only generated inside the mesophyll objects but not in their membranes. The Markers image is illustrated in \figurename~\ref{fig_seg}B. 

The marked watershed distance transform \cite{soille90} is then used to segment the Tissue density image using the Markers image. The resulting watershed dams are added to the Markers image as membrane markers. A second watershed transform is then used with the updated markers resulting in a final segmentation into separated mesophyll objects and membranes. 

\subsubsection{Tissue structure}
In order to extract the tissue structure, the skeleton of the membranes is build using scikit-image \textit{skeletonize} function based on \cite{zhang84} and followed by pruning to remove the 'barbules'. The skeleton is represented by a blue line in \figurename~\ref{fig_seg}C. The multiple points, i.e., the border pixels having at least three connected neighbors, are then detected using classical hit-or-miss transform. Each multiple point is associated to its neighboring objects. The multiple (mostly triple) points are represented as red dots in \figurename~\ref{fig_seg}C. For each object, a straight contour is build by joining the pairs of adjacent multiple points by a line segment (\figurename~\ref{fig_seg}D). These segments are used for calculating the specific features defined below.

\subsubsection{Vein detection}

Most objects are either cells or voids occurring between adjacent cells, but some group of smaller objects are actually veins of the vascular system.
In order to prevent confusion due to the similar size of the smallest voids and the cells forming the veins, we specifically detect the latter using prior knowledge about their structure. Veins are formed by small aggregate of small cells whereas voids are usually surrounded by larger cells. Veins are detected by finding clusters of adjacent objects having a small area (i.e. smaller than a threshold value arbitrarily set to 1000\,pixels\textsuperscript{2}) and forming groups of a minimum number of objects (here set to ten after visual examination of the slide). The isolated small objects are ignored. \figurename~\ref{fig_leaf_morphology} illustrates detected veins with a purple overlay.
We also ignore objects in contact with the tile borders, since their shape features may be altered; the number of objects lost has been estimated to be less than $4\%$ of the total number of available objects.

\subsubsection{Object labeling}

The low resolution Label image and the Epithelium mask are cropped to the corresponding tile region and resampled at the same resolution as the tile.
All the segmented objects receive their corresponding leave piece label from the Label image.
The objects having their centroid inside the epithelium mask are tagged as 'epithelium cells' (green part in \figurename~\ref{fig_leaf_morphology}). 
The objects belonging to vein clusters are tagged as 'vein' (purple clusters in \figurename~\ref{fig_leaf_morphology}) and objects with a centroid inside the Phenolic mask are tagged 'phenolic' (brown overlay in \figurename~\ref{fig_leaf_morphology}). The remaining untagged objects are further identified by the supervised classification process into cell objects and void objects.

%===============================
\subsection{Feature extraction}

Careful microscopic examination of tissue shows that air lacunae or voids are deflated due the convex surface of neighboring cells, which are inflated (labelled 'V' and 'C' respectively in \figurename~\ref{fig_features}A). Convexity can be quantified by means of conventional object descriptors \cite{Sonka07}, but they are not sufficiently discriminating to reliably detect voids. Indeed, a cell in contact with a void typically exhibits convexity due to its internal pressure. However, some concave parts may appear due to a segmentation artifact. Indeed, the membrane thickness is close to the scanner resolution and cell corners may be difficult to locate accurately.
Likewise, few voids have parts of their contour that seem inflated due to imperfect segmentation, e.g., as a result of presence of 'debris' inside the void.
In these different cases, conventional convexity features based on circularity or the strict convex hull extraction, such the ones available in the used library, are of limited use. Specific descriptors were therefore developed to enrich object description, as described below.

\subsubsection{Generic features}
We investigated conventional features known to discriminate objects with different shapes in terms of convexity. For example, Pieczywek et al. \cite{pieczywek12} uses circularity and shape roughness for separating cells from inter-cellular spaces in apple tissue. In the present study, for each segmented object we extract the following set of classical shape descriptors (provided by scikit-image python image processing library \cite{vanderwalt14}): $area$, $convex\, area$, $eccentricity$, $equivalent\, diameter$, $extent$, $major\, axis\, length$, $minor \,axis\, length$, $perimeter$ and $solidity$ (i.e., the ratio of the region area to the convex hull area).

\subsubsection{Specific features.}
\begin{figure}[ht]
\centering
\begin{tabular}{rr}
   A. \includegraphics[width=2.0in]{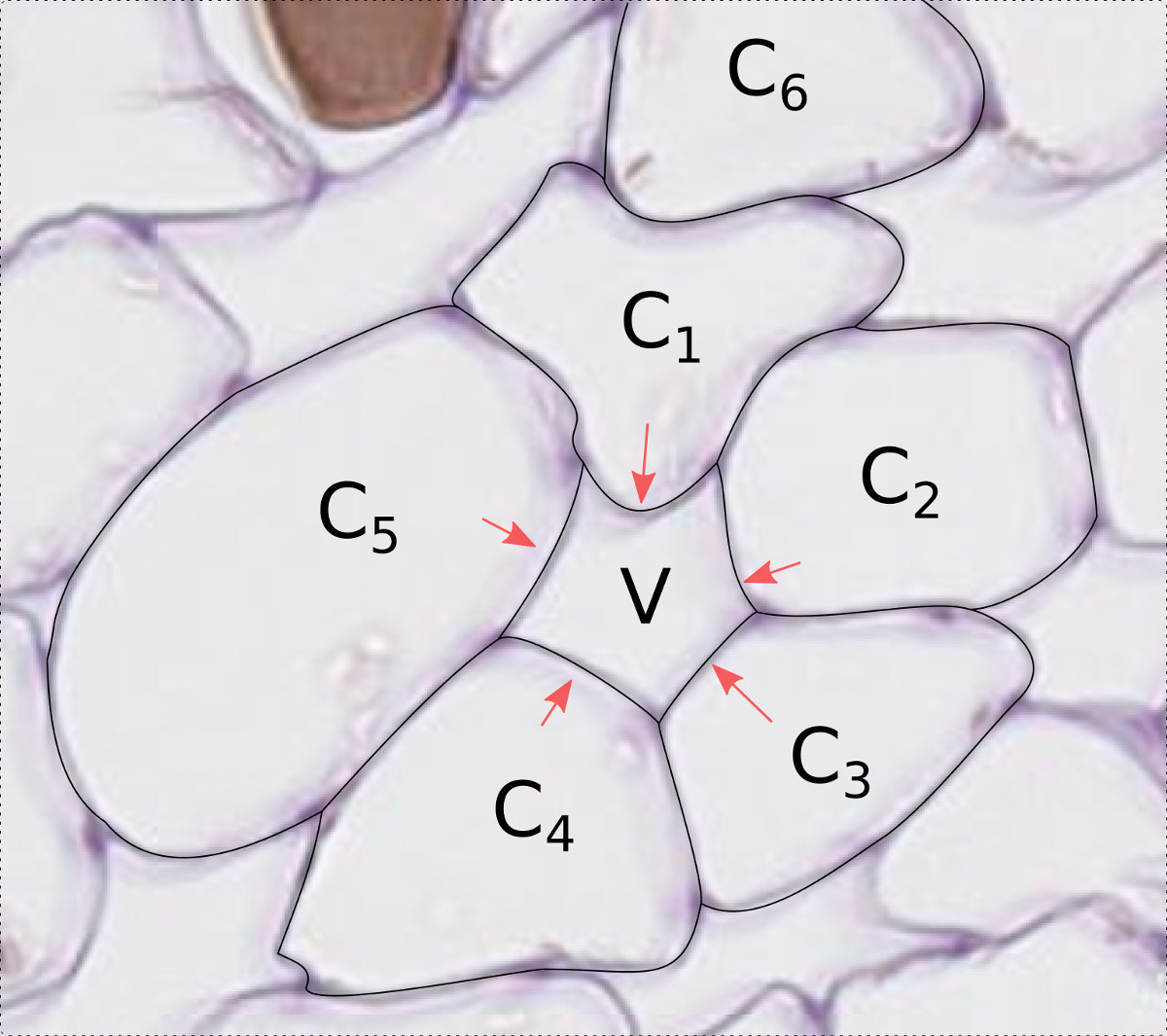} & B. \includegraphics[width=2.0in]{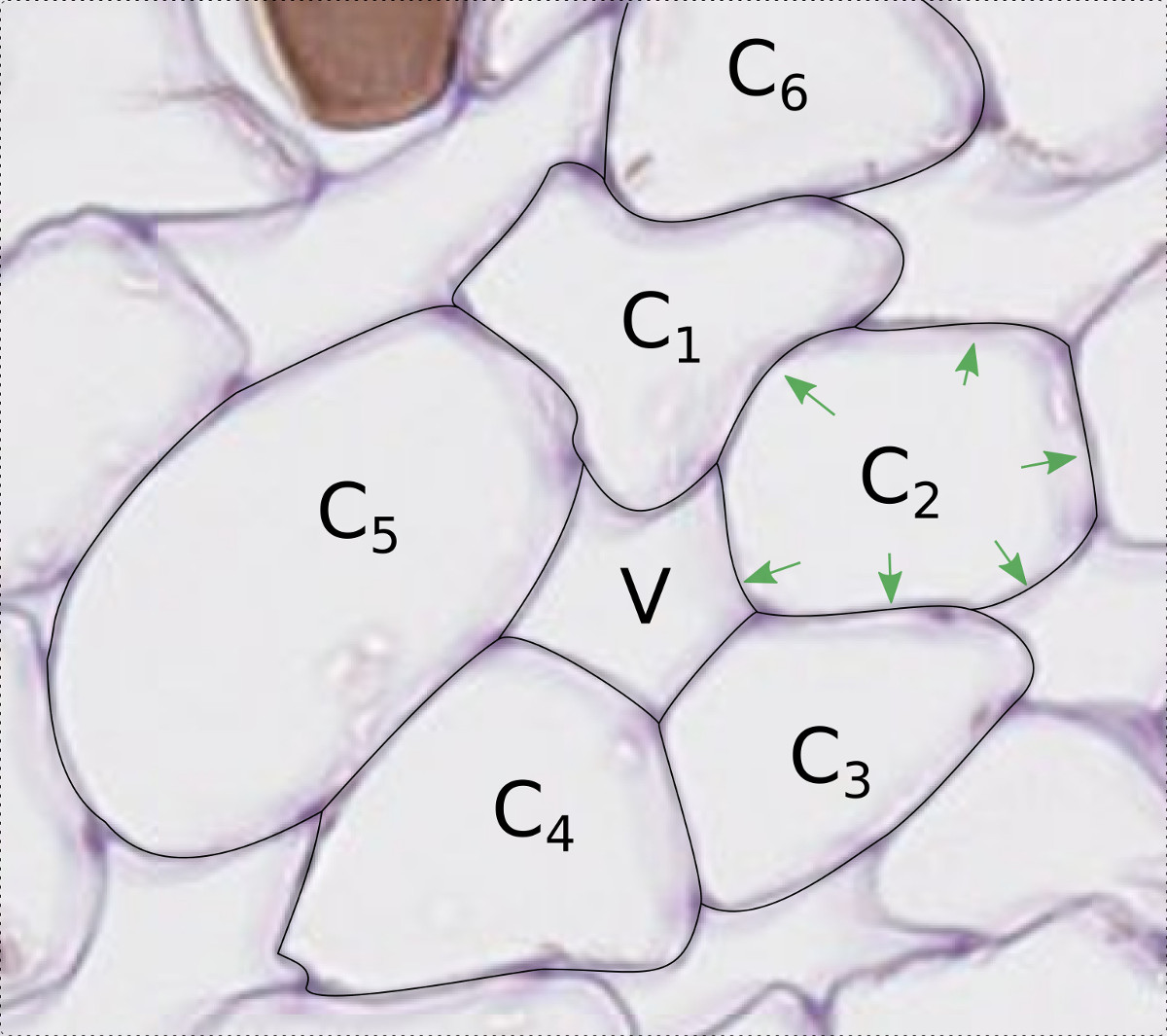}\\
   C. \includegraphics[width=2.0in]{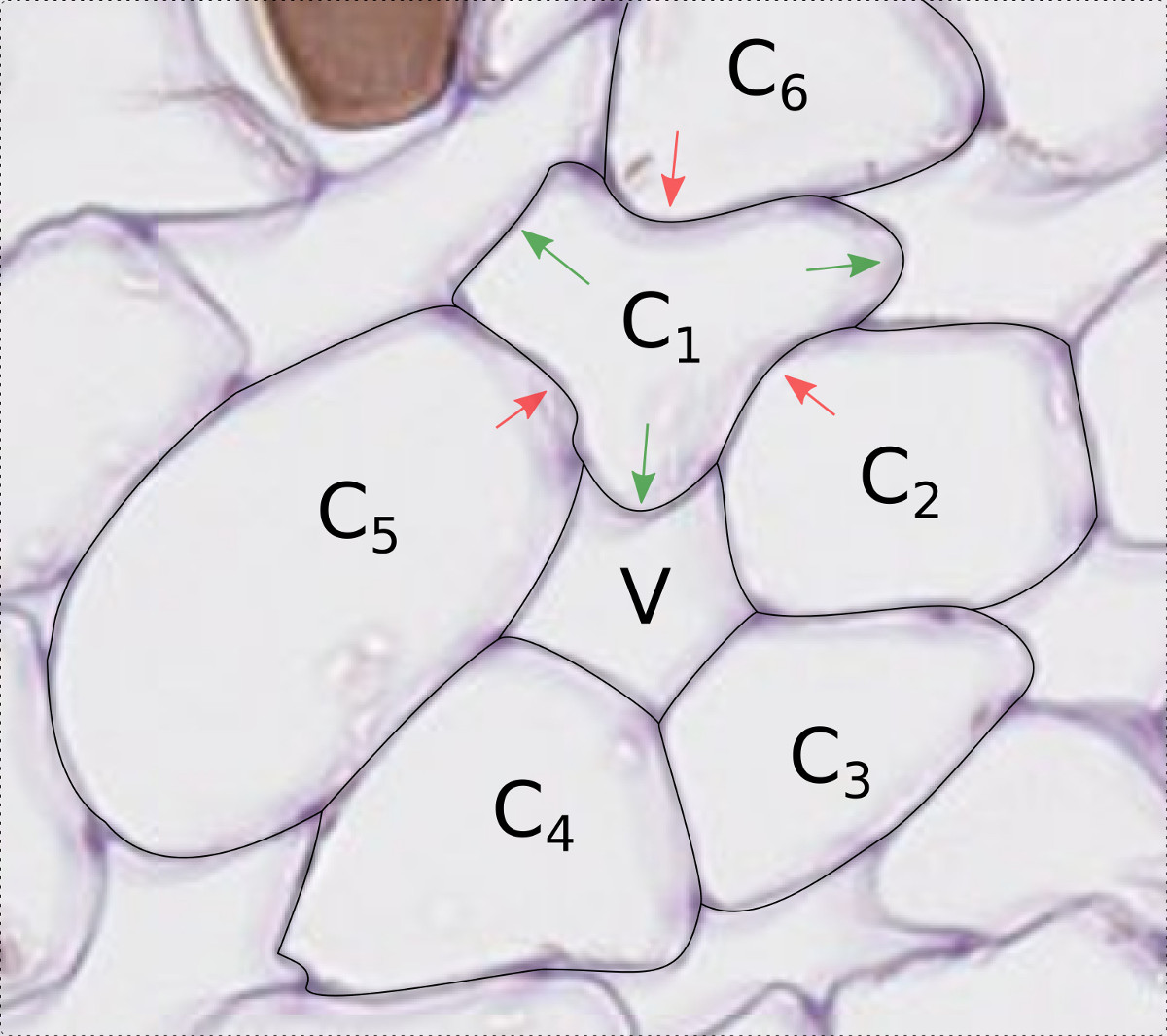}  & D. \includegraphics[height=1.76in]{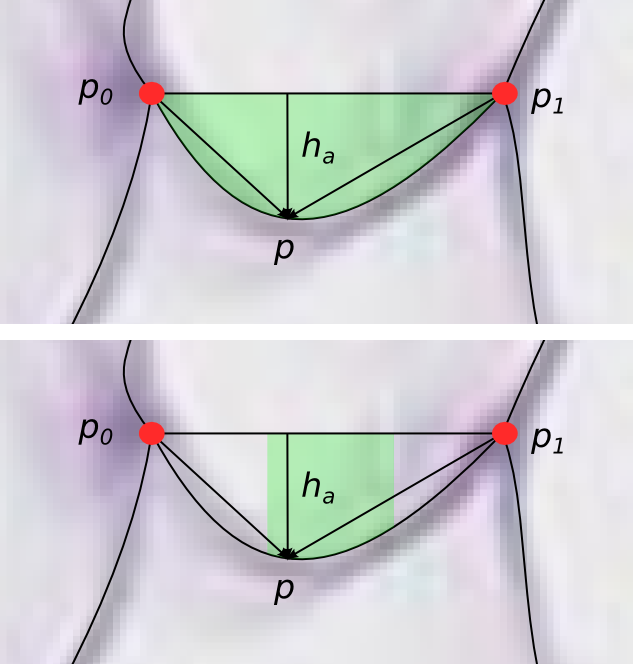}\\
\end{tabular}
\caption{\textbf{Specific Features.}
A: example of cells ($C_i$) around a void ($V$) where arrows show the deflated aspect of the void object, B: in contrast, cells are inflated, C: example of a more complex object having both inflated and deflated portions, D: definition of the membrane inflation ($h_a$), measures are summed on the complete segment $p_0$-$p_1$  or on its central part only (green area) to be less sensitive to segmentation noise.
}
\label{fig_features}
\end{figure}

Due to the limited effectiveness of conventional descriptors (see results), new ones were developed to describe aerenchyma tissue, specifically to identify air lacunae surrounded by mesophyll cells.
To characterize concave and convex object borders, the segments joining two consecutive multiple points of the skeleton are computed. As illustrated in \figurename~\ref{fig_features}D, the signed distance $h$ between the straight line of the skeleton joining two consecutive multiple points (labeled $\mathbf{p}_0$ and $\mathbf{p}_1$) and each pixel, $\mathbf{p}_i$, of the actual border line is calculated as
\begin{equation}
\label{eq:h}
    h(i) = 2\frac{(\mathbf{p}_0-\mathbf{p}_i) \times ( \mathbf{p}_1-\mathbf{p}_i)}{d}
    \;\mathrm{  with }\;d = ||\mathbf{p}_1-\mathbf{p}_0||
\end{equation}
A negative value of $h$ corresponds to an deflated object border, i.e., a void.

Distances $h$ are then summed on the $n$ pixels of each object segment using:
\begin{equation}
\label{eq:s}
    area = \sum_{i = 0 \dots n } h(i),
\begin{cases}
    area < -15  & \text{the segment is considered as inflated}\\
    area > 40  & \text{the segment is considered as deflated}\\
    \text{else undefined}
\end{cases}
\end{equation}

When the membrane is inflated (e.g., $h_a$ in \figurename~\ref{fig_features}), most of the actual path is located outside the polygon. Conversely, if the membrane is deflated (e.g., $h_b$ in \figurename~\ref{fig_features}), a significant part of it is inside the object polygon. 

Let define $n_{seg}$ as the number of boundary segments of an object, e.g., in \figurename~\ref{fig_features} $n_{seg}=6$ for object C and $n_{seg}=5$ for object V. Similarly, $n_{out}$ and $n_{in}$ are the numbers of inflated and deflated segments, respectively.

$area_{out}$ and $area_{in}$ are the inflated and deflated areas for the complete object, respectively:

\begin{eqnarray}
\label{eq:area_in_out}
    area_{in} = \sum_{i=0\dots N} 
\begin{cases}
    +h(i) & h(i) > 0 \\
    0 & h(i) < 0
\end{cases}\\
    area_{out} = \sum_{i=0\dots N}
\begin{cases}
    0 & h(i) > 0 \\
    -h(i) & h(i) < 0
\end{cases}
\end{eqnarray}
where the $h$ distances are summed along each segment of the object, i.e. on a total of $N$ pixels.

To limit the small variations in position of the triple-points due to eventual segmentation issue, we also develop another version of these features for which only the central part of each segment (i.e. the central third of the pixels of the segment) is taken into account for the measure (green area in \figurename~\ref{fig_features}D), as defined in Eq.~\ref{eq:areac}.

\begin{equation}
\label{eq:areac}
    area_c = \sum_{i = \lfloor n/3 \rfloor \dots \lfloor 2n/3\rfloor } h(i),\text{ with }
\begin{cases}
    area_c < -40 & \text{the segment is considered as inflated}\\
    area_c > 40 & \text{the segment is considered as deflated}\\
    \text{else undefined}
\end{cases}
\end{equation}

The above threshold values defining inflation or deflation were determined by trial and error.
The corresponding features are $n_{cout}$, $n_{cin}$, $area_{cout}$ and $area_{cin}$, where $c$ stands for centered.

\subsubsection{Discriminant ability}
The discriminant ability of each feature with respect to the two-class "cell vs. void" problem was evaluated using the Mann-Whitney statistics ($U$) which compares two independent groups of quantitative data. It was implemented using the function \textit{mannwhitneyu()} from \textit{scipy} library \cite{scipy}. Because of the large number of available cases in each group, the p-values is not really informative \cite{sullivan12}. In such situation, a little shift between the two distributions is significant. We therefore compute a quality estimate $Q$ for the $U$ statistics, knowing that the maximum $U$ value (as well as the sum of the $U$ values computed on the two data groups) is the product of the two sample sizes ($n_0,n_1$): 

\begin{eqnarray}
\label{eq:quality}
Q = \max (\frac{U}{n_0n_1},\frac{n_0n_1-U}{n_0n_1})
\end{eqnarray}

This expression allows us not to worry about which data group (i.e. the smallest or the largest) $U$ has been calculated on. The $Q$ values range from 0.5 to 1, depending on the extent of overlap between the two groups of data to be distinguished, i.e. from full to no distribution overlap, respectively. The closer $Q$ is to 1 for a feature, the more discriminating this feature is.

\subsection{Supervised classification}

\subsubsection{Supervision and datasets}\label{supervision}
Each tile was visually examined by an expert who identified and filled each void by means of an image processing software (here, Adobe Photoshop), the remaining objects (except epithelium and veins) being considered as cells. For the slide investigated, a total of 5149 voids were annotated. Since the examination time was purposely limited, the supervision is not perfect: some voids were missed and, for the less contrasted parts of the images, some cells may have been identified as a void. For each section, the total number of pixels corresponding to cell and void were counted to estimate the respective areas and void fraction per section. 

From the complete supervised dataset composed of 55940~cells and 5149~voids, all objects of one tile were removed to be used as an independent test set (1383 cells and 247 voids); all objects already tagged as vein or epithelium are also removed from the data (10051 objects). In order to limit the training time and balance the training set, one tenth of the cell cases were randomly subsampled. The final training set is composed of 4902 void samples and an equal number of cell samples.

\subsubsection{Classifier}
We used \textit{SVM}, which is based on \textit{libsvm} \cite{chang11}, one of the state-of-the-art machine learning-based approaches. The algorithm is provided by the \textit{scikit-learn} python library \cite{pedregosa11}. A cross-validation-based grid search method (applied on the training set only) is used to estimate the best classifier parameters. The linear and $\mathrm{RBF}$ kernels were tested.

After determining the best parameters for the classifier, the model was retrained on the whole training data and then used to classify the independent test set to assess the performances.

Concerning the input features, the 9 generic features ($area$, $convex\, area$,  $eccentricity$, $equivalent\,diameter$, $extent$, $major\, axis\, length$, $minor\, axis\, length$, $perimeter$, $solidity$) are first considered and then enriched by the 9 specific features ($n_{in}$, $n_{cin}$, $n_{out}$, $n_{cout}$, $n_{seg}$, $area_{in}$, $area_{cin}$, $area_{out}$, $area_{cout}$). 

All feature are normalized using the \textit{sklearn RobustScaler} method that scales the data between percentile .1 and percentile .99 of the training set.

The McNemar's test \cite{mcnemar47} is used to compare the classifier performances obtained on the independent test set using the \textit{mcnemar} function of the \textit{statsmodels} library (git://github.com/statsmodels/statsmodels.git).
%=================

\begin{table}[ht]
\centering
\caption{\textbf{Discriminant ability of the extracted object features.}}
\input{table_feat.tex}
\begin{flushleft} Results of the Mann-Whitney test computed between the voids (n=4920) and the cells (n=45889) from the complete supervised dataset. The features are sorted according to their $Q$ value computed as defined in~(\ref{eq:quality}). $Q$ takes values between 0.5 and 1, from complete to no overlap between the two data distributions. 
\end{flushleft}
\label{table_feat_sorted}
\end{table}

\section{Results}

\subsection{Discriminatory ability of the features}

\begin{figure}[ht]
\centering
\begin{tabular}{cc}
   A. \includegraphics[width=2.0in]{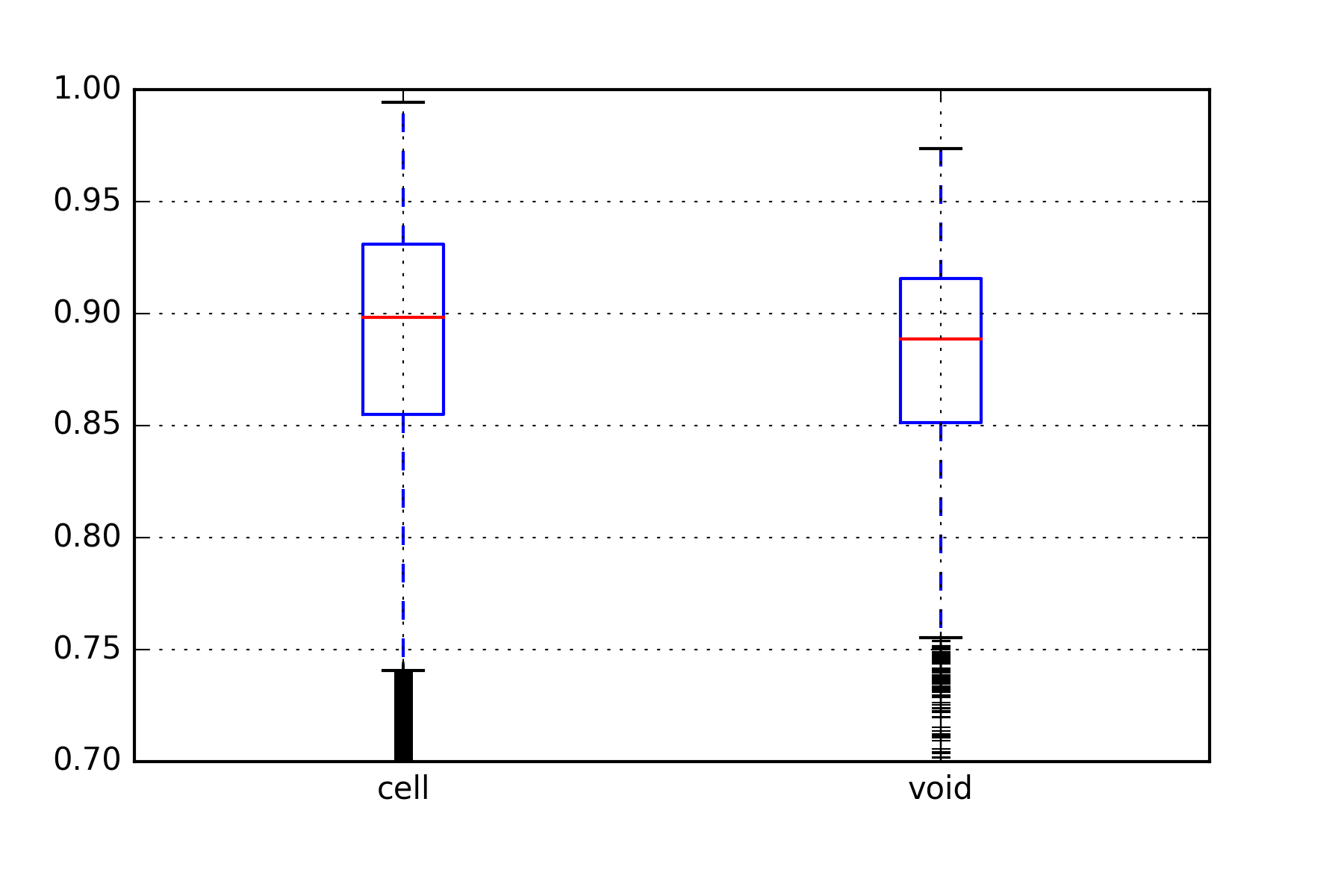} & B. \includegraphics[width=2.0in]{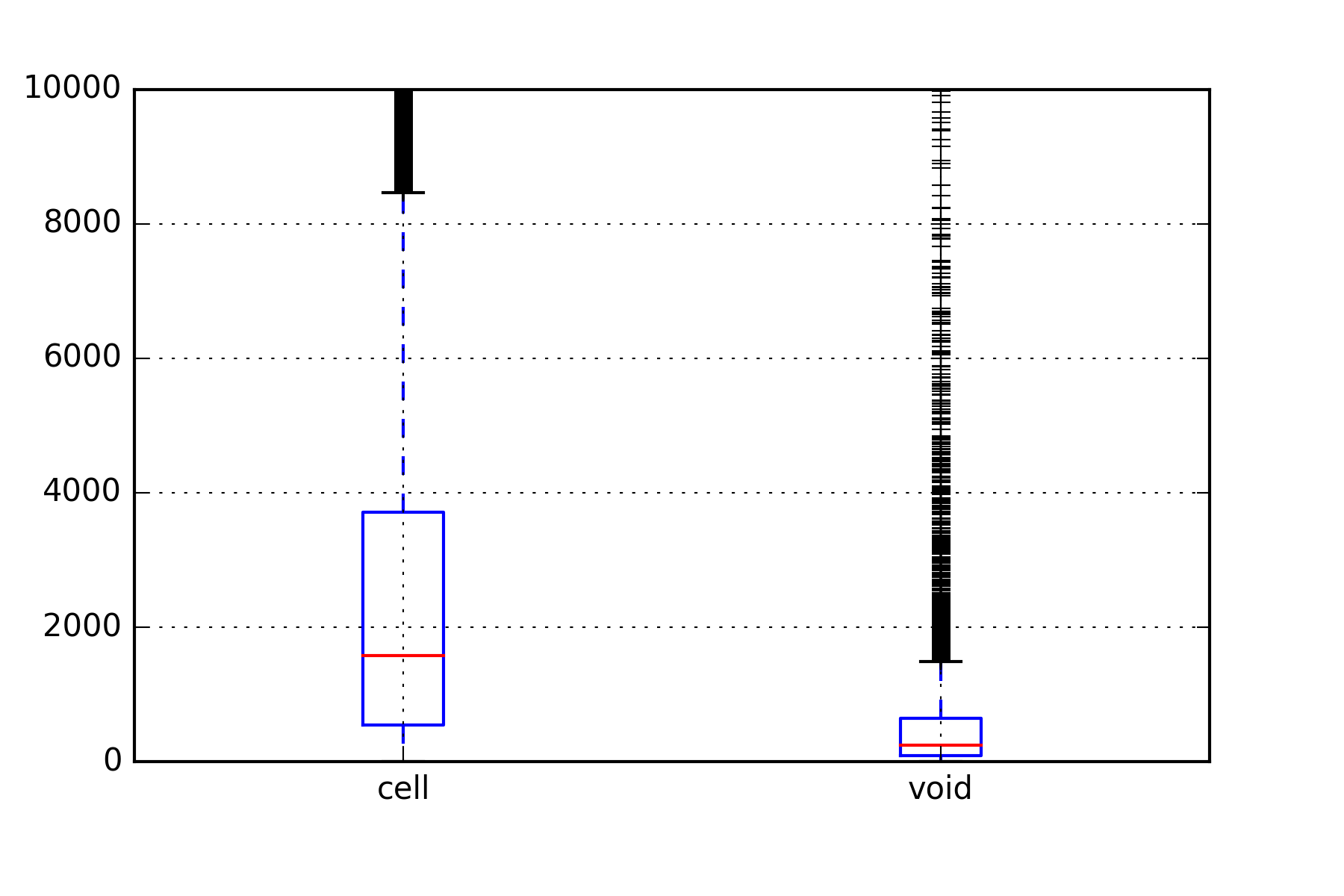}\\
   C. \includegraphics[width=2.0in]{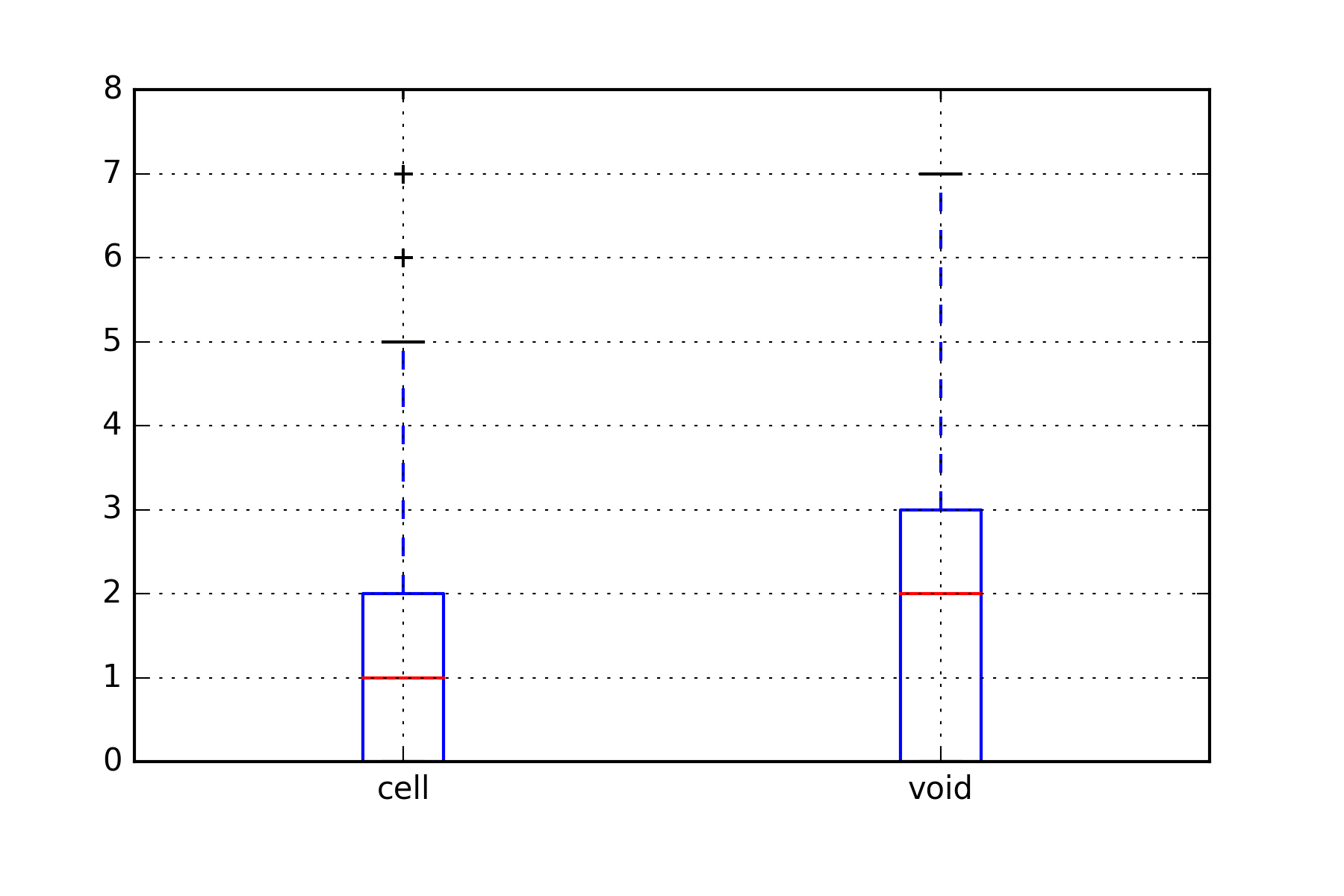} & D. \includegraphics[width=2.0in]{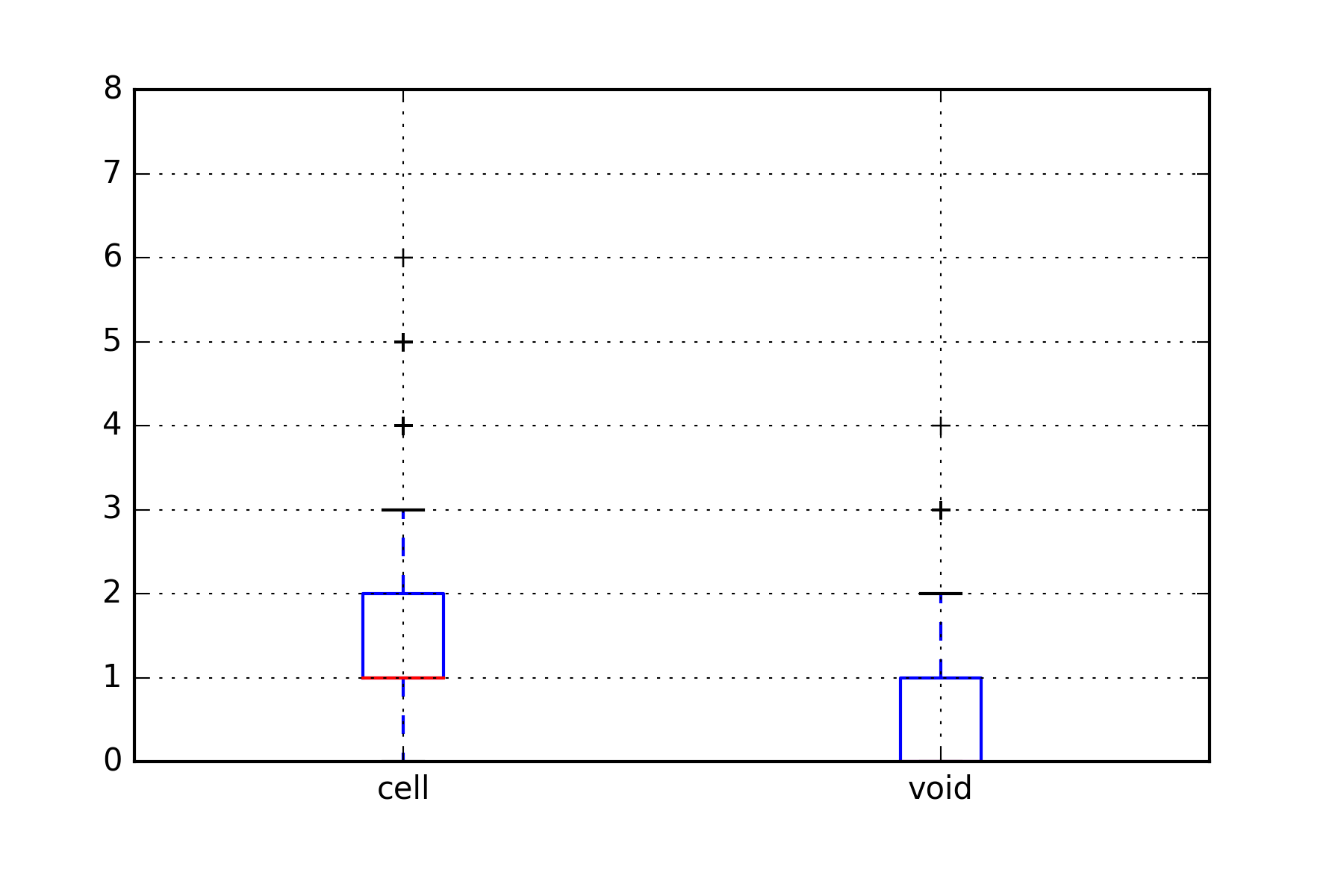}\\
  
\end{tabular}
\caption{{\bf Feature boxplots between cells and voids.} A: generic solidity; B: inflated surface of object (\micro\squaremetre), C and D: number of deflated and inflated boundaries per object. The box extends from the lower to upper quartile values with the red line being the median. The whiskers extend from the non-outlier minimum to non-outlier maximum, while crosses are outlier values.}
\label{fig_solidity}
\end{figure}

The generic features describe satisfactorily the basic shape of the objects but miss key characteristics. Contrary to the results obtained in \cite{pieczywek12} where cells are almost exclusively convex, the discrimination obtained with the generic features was ineffective for the present purpose.
Detailed analysis of the data show that this is mainly due to the facts that cells and voids share a same range of value in terms of area and that the cell and void objects are distributed on a relatively regular grid making their convexity similar. In addition, the cells and voids have largely overlapping distributions of solidity values, as illustrated in \figurename~\ref{fig_solidity}A. In contrast, the new specific features exhibit a higher discriminating ability, as illustrated in
\figurename~\ref{fig_solidity}C and~\ref{fig_solidity}D.
Voids exhibit more deflated segment than cells, similarly the inflated surface is larger for cells than for voids (\figurename~\ref{fig_solidity}B). 

The $Q$ values in Table~\ref{table_feat_sorted} confirm all these observations and evidences that new features, especially those characterizing inflated properties (i.e. with the $ _{out}$ index), exhibit a higher discriminating ability than the others. Of the generic features, only $extend$ and $solidity$ exhibit significant differences but with low $Q$ values, i.e. near 0.50.

\subsection{Classifier performance}

Whatever the input feature set used, the cross-validation-based grid search method performed on the training set determine the same set of parameters for the sklearn SVC method ($kernel = \mathrm{RBF}$, $\gamma = 1000/\#feat$, $C = .01$). Using these parameters, each SVM model is retrained on the whole training data and then applied on the independent test set. Tables \ref{table_svm_classic} and \ref{table_svm_all}  detail the classification results so obtained using either the 9 generic features only or the complete set of 18 features, respectively. In view of the results provided in Table~\ref{table_feat_sorted}, a third model is trained using the 11 significant features as input, i.e. the specific features completed by $extend$ and $solidity$. The results are provided in Table \ref{table_svm_ext}.

In Table \ref{table_svm_all}  (all features) the correct classification rates for cell and void objects are respectively $80.2\%$ and $90.4\%$, which are better than the results obtained with the generic features only (i.e. $72.1\%$ and $80.8\%$, Table \ref{table_svm_classic}). It should be noted that the classification rates computed in terms of object area is better for the cells (increasing to $86.4\%$ in Table \ref{table_svm_all}) but remaining similar for the voids ($91.8\%$ in Table \ref{table_svm_all}), showing that the misclassified cells are small objects. 
The results obtained with only the significant features fall between the other two (Table \ref{table_svm_ext}: $76.0\%$ and $87.0\%$ for cell and void object classification). 

\begin{table}[ht]
\centering
\caption{Test set classification results: SVM classifier using the generic features only (9).}
\input{table_pred_classic.tex}
\begin{flushleft} 
\end{flushleft}
\label{table_svm_classic}
\end{table}

\begin{table}[ht]
\centering
\caption{Test set classification results: SVM classifier using all available features (18).}
\input{table_pred_all.tex}
\begin{flushleft} 
\end{flushleft}
\label{table_svm_all}
\end{table}

\begin{table}[ht]
\centering
\caption{Test set classification results: SVM classifier using all significant features (11).}
\input{table_pred_ext.tex}
\begin{flushleft} 
\end{flushleft}
\label{table_svm_ext}
\end{table}

\begin{table}[ht]
\centering
\caption{McNemar comparison of the object predictions obtained using the three feature sets}
\input{table_p-value.tex}
\begin{flushleft} 
\end{flushleft}
\label{table_svm_pvalue}
\end{table}

In Table \ref{table_svm_pvalue} the McNemar's test results evidence that the SVM model using all the 18 features as input is significantly better than the two others and that the model using the significant features is significantly better than the generic feature-based model, but with a lower level of significance.

\begin{figure}[ht]
\centering
\includegraphics[width=4.0in]{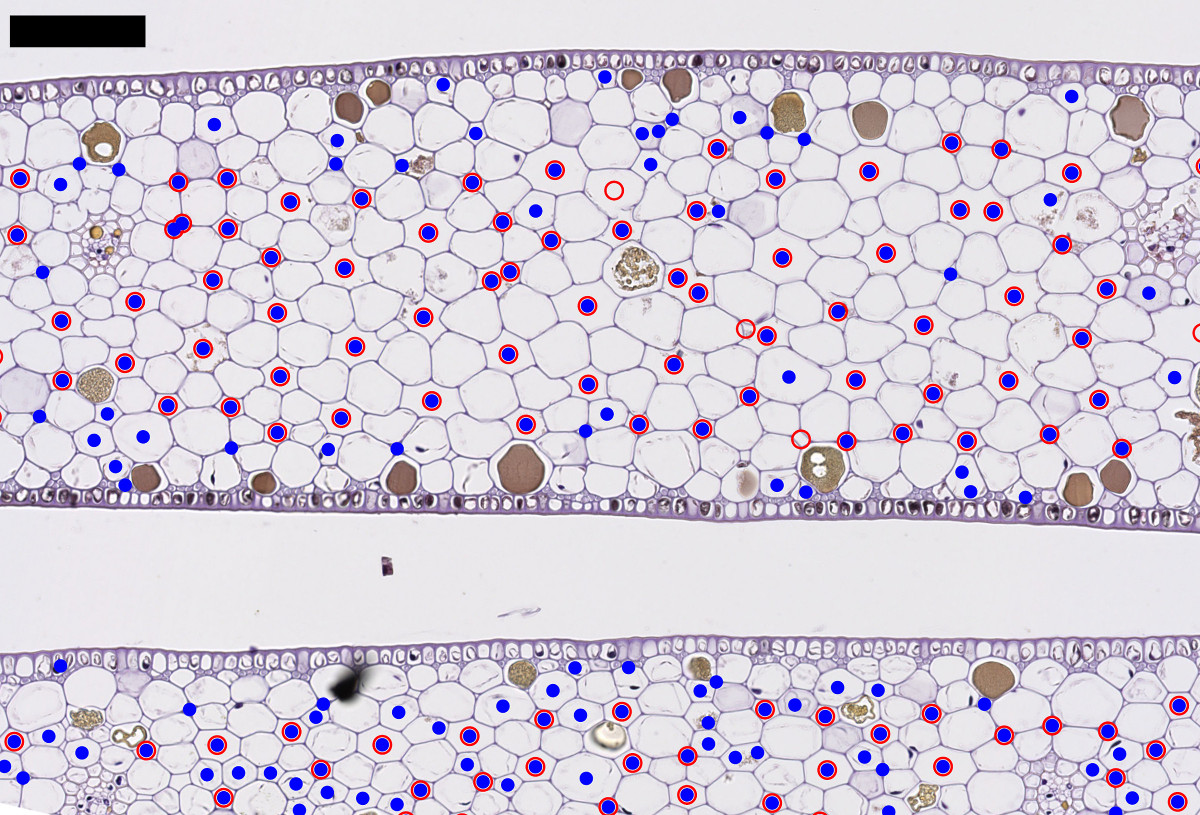}
\caption{\textbf{Cell vs void classification.}
Prediction on an independent data set. Red circles and blue dots correspond to the expert labeling and the SVM predictions obtained with the 18 features as input, respectively (scale bar = 100\micro\meter).
}
\label{fig_pred_tile}
\end{figure}

However, in terms of predictive value, the void label is correctly predicted by the best classifier for $40.3\%$ of the objects only, representing $50.3\%$ of the pixels (Table \ref{table_svm_all}). \figurename~\ref{fig_pred_tile} illustrates these results by comparing the so predicted voids (blue dots) to the expert labeling (red circles). A careful examination of the false positive objects (i.e. blue dots without red circle in figurename~\ref{fig_pred_tile}) actually reveals unlabelled voids, as discussed below.

\subsection{Seagrass tissue morphology and anatomy characterization}

In order to extract morphology and anatomy characteristics, image is processed as described before, epithelium and vein are detected and the remaining objects are classified between cell and void using the best SVM classifier. To study the progression of these anatomical characteristics, each object/structure is associated to its blade segment, which in turn is localized in a specific part of the leave (base, middle or apex).

\subsubsection{Vein number and size distribution}

\figurename~\ref{fig_vein} illustrates various measures obtained for the leaf vascular system. \figurename~\ref{fig_vein}A shows that the number of veins is lower at the apex, decreasing progressively along the leaf blade.  \figurename~\ref{fig_vein}B shows that the total surface occupied by the veins decreases from base to apex while \figurename~\ref{fig_vein}C. shows that the vein equivalent diameter decreases from base to apex. It is noted that since a few vein objects (approx. 20\%) were split between tiles their size distribution was slightly altered without justifying a special treatment. However, the total surface area of vein is unbiased.

\subsubsection{Epidermis}
\begin{figure}[ht]
\centering
\begin{tabular}{cc}
   A. \includegraphics[width=2.0in]{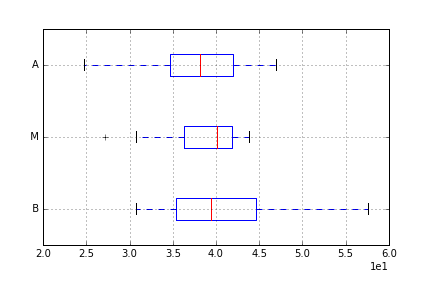} & B. \includegraphics[width=2.0in]{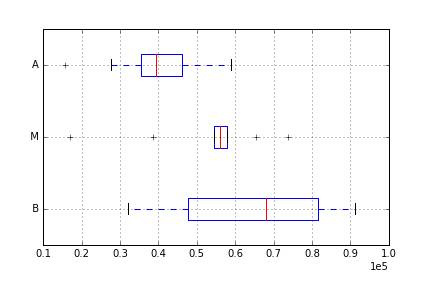}\\
   C. \includegraphics[width=2.0in]{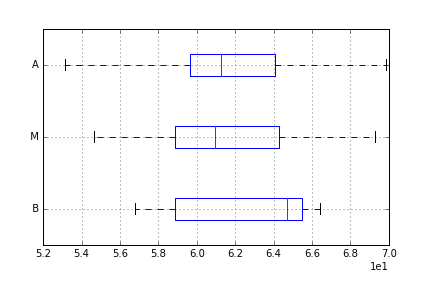} & D. \includegraphics[width=2.0in]{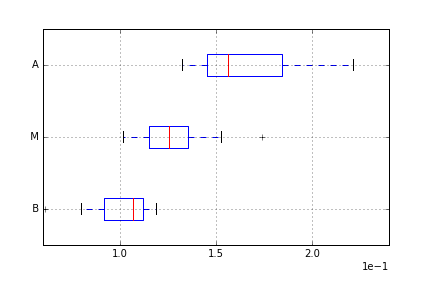}\\
  
\end{tabular}
\caption{{\bf Vein data distribution.} A. Mean vessel number count per section, B. Vessel cell area sum per section (\micro\squaremetre), C. Mean vessel equivalent diameter (\micro\meter), D. Fraction of the epithelium per section. The box-and-whisker plot depicts the descriptive statistics evaluated from the corresponding cross-sections. For boxplot interpretation cf. Figure \ref{fig_solidity}.
}
\label{fig_vein}
\end{figure}

Another measure can be inferred such as the fraction of epidermal cells with respect to the total section area. \figurename~\ref{fig_vein}D. shows that this ratio increases from base to apex, which reflects the relative uniformity of the epidermis thickness along the leaf while the total tissue section is increasing.

\subsubsection{Void fraction}
As vascular land plants, the void fraction of seagrasses can be defined as the ratio of the void volume and total volume of tissue (composed of epithelium, veins and mesophyll cells):

\begin{eqnarray}
\label{eq:void_frac}
VF = \frac{V_{void}}{V_{total}}.
\end{eqnarray}

\begin{figure}[ht]
\centering
\begin{tabular}{cc}
   A. \includegraphics[width=2.0in]{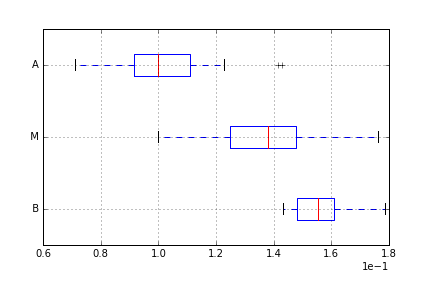} & B. \includegraphics[width=2.0in]{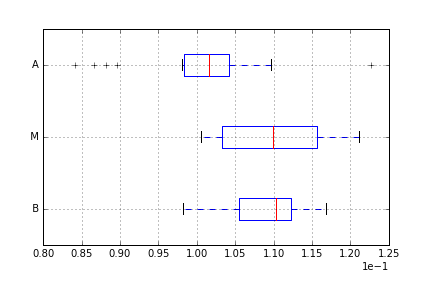}\\
   C. \includegraphics[width=2.0in]{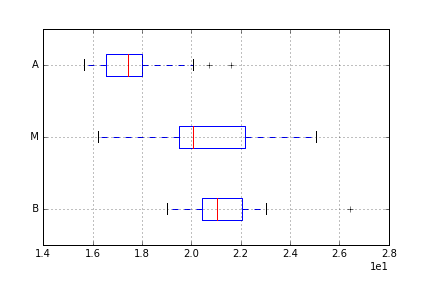} & D. \includegraphics[width=2.0in]{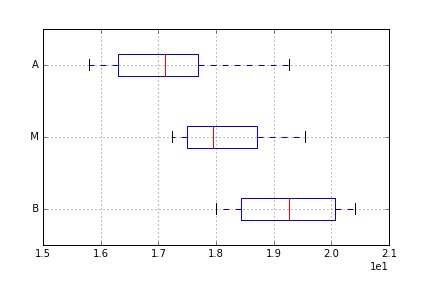}\\
\end{tabular}

\caption{\textbf{Void fraction at the base (B), middle (M) and apex (A) of the leaf blade.}:A. Section Void fraction, B. Membrane ratio, C. Cell median equiv diameter (\micro\squaremetre), D. Void median equiv diameter (\micro\squaremetre). The box-and-whisker plot depicts the descriptive statistics evaluated from the corresponding cross-sections, as in Fig.~\ref{fig_vein}. 
}
\label{fig_boxplots}
\end{figure}

Because of a poor classifier predictive value, the amount of false positive pixels is too high to allow a useful void/tissue fraction estimation. However, if we make the hypothesis (not yet verified) that the error made is constant along the leave, we can still have an idea of the void/tissue fraction evolution along the leave as illustrated in \figurename~\ref{fig_boxplots}. \figurename~\ref{fig_boxplots}A summarizes the distributions of the void fraction values obtained after grouping the blade segments per location (i.e. base, middle and apex, including approximately 10 segments each). The median decreases from $15\%$ at the basal end to $10\%$ at the apex end. 

\section{Discussion and conclusion}

A method is proposed to characterize the leaf-blade aerenchyma of \emph{P. oceanica} seagrass by means of WSI, as commonly used for human or animal tissue. As discussed by Visser et al. \cite{visser03} for land plants, microscopy gives better insight into aerenchyma structure than buoyancy-based approaches but is subject to large uncertainty when applied to a too limited number of cross sections. Our approach is able to solve the issue by increasing the number of sections thanks to automatic analysis of a series of whole images of thin slices cut from paraffin blocks embedding stacks of short blade segments. Microtomy enables a resolution of \unit{4} to 5{\micro\metre} in the blade axial direction. Successive sections would thus enable reconstruction of the aerenchyma structure of a whole leaf blade with transverse and axial resolutions respectively higher than and comparable to X-ray microtomography. Most importantly, our approach provides the tissue composition at the cellular level, an information that is not accessible to X-ray absorption-based high-resolution micro-CT due to low density contrast between tissue components except void and vascular when taken as a whole.

Based on WSI, image processing algorithms were developed that identify cells from epidermis, mesophyll, vascular system and the air lacunae. This approach gives both statistical measures and microscopic insight of the tissue morphology. WSI allows also a topological description of the objects of interest at the microscopic level. The same image can provide information about the number of cells surrounding a void, the average distance between voids, veins position and number, etc. %The WSI acquisition provides the anatomy of the tissue but also 
This detailed information cannot be provided by X-ray absorption alone due to the faint density contrast between the different cell types. However, the dehydration procedure needed for embedding tissue in resin or paraffin may lead to artifacts affecting cell and void size. Since, by essence, the method is sensitive to microscopic features, its performance depends on the quality of tissue slide which can be altered by scratches, cuts, folding or desiccation.

Visser et al. \cite{visser03} also mentioned that some plant species exhibit very similar pattern for cell and lacunae. We observe the same feature for the investigated seagrass species, in particular for the \emph{P. oceanica}, where generic features such those described in Pieczywek et al. \cite{pieczywek12} lack the level of details required to distinguish voids from cells. In the present pilot study, we propose specialized shape descriptors which significantly improve classification results to discriminate voids from cells on HE-stained tissue slices. 

However, the classification approach lacks to provide a sufficiently specific prediction, leading to a high false positive rate for void objects. By a close examination, it turns out that many small objects that are falsely classified as void (with respect to supervision) actually are unlabelled voids. This situation occurs mainly because the supervision has been made only on the void class, making the hypothesis that all the unlabeled objects are cells. Improving supervision for the small objects is required to improve prediction results. An alternative is to develop other learning strategies taking these supervision defects into account \cite{jiang17,foucart19}. 

Examination of more cross sections, WSI and routine microscope slides (not shown), suggests that typically \emph{P. oceanica} voids are regularly spaced and surrounded by an average of six cells. However, some variations of this pattern were observed where the voids in the base section tend to have more neighboring cells than six. This topological/graph aspect might also be used as post-processing to detect eventual bad cell/void classifications.

The provided plant section description provides absolute area and area ratio for the different tissue types, and also gives information about the topology of the section, i.e. the relative distribution of the different tissue types and their connectivity. 
In addition, preliminary tests show that the low level pixel approach should also enable the detection of the fiber cells which are not only of main interest for the plant acoustic structure but also very difficult to detect with other state-of-the art imaging modalities (data not shown). 

While the present approach is based on standard 2D sections of the plant, the approach allows to efficiently multiply the number of samples along the leave, which in turn gives a dense description of all the parameters along the leave. This basic research contributes to enhance acoustic mapping of these marine habitats and to develop acoustic sensing to assess the health of these world’s most valuable but fragile ecosystems, a marker of the health of the larger estuarine and marine ecosystems \cite{boudouresque06,diaz08}.

Although the described method was applied specifically to \emph{P. oceanica}, it appears that the void vs. cell recognition is relatively challenging in this case compared to other species where voids can have a very specific aspect \cite{visser03}. The same methodology could be therefore applied to other species.

\section*{Acknowledgments}
The authors acknowledge The Office of Naval Research and the Fonds d'Encouragement à la Recherche (FER)-French Community of Belgium for their generous support of this research.
The Centre for Microscopy and Molecular Imaging (CMMI) is supported by the European Regional Development Fund and the Walloon Region.
The authors thank Mélanie Derock (DIAPath, CMMI) for her technical assistance, as well as Nikos Giannenakis, Louise Hermand, Fanny Hermand and Arthur Piolet for their volunteer participation in the field and laboratory work. The expertise of the Laboratoire de Biotechnologie Végétale (LBV) of the Université libre de Bruxelles (ULB) is also appreciated.

\bibliographystyle{unsrt}  
\bibliography{refs.bib}

\section*{Supporting Information}

\paragraph*{WSI file}
\label{S1_zenodo}
The original whole slide image used to produce the results for this paper is available in NDPI format at https://zenodo.org/record/46395/.

\end{document}

%% file: table_feat.tex
\begin{tabular}{l|ll}
                    & MW-pvalue & $Q$   \\ \hline
$Area$                & 4.62E-01  & 0.50        \\
$Equivalent\,  diameter$ & 4.62E-01  & 0.50         \\
%$Filled\,  area$         & 4.62E-01  & 0.50         \\
$Perimeter$           & 4.15E-01  & 0.50      \\
$Convex\, area$         & 3.36E-01  & 0.50          \\
$Major\, axis\,  length$   & 2.18E-01  & 0.50        \\
$Minor\,  axis\,  length$   & 1.70E-01  & 0.50          \\
$Eccentricity$        & 1.45E-01  & 0.50        \\
$Extent$              & 3.59E-37  & 0.55          \\
$Solidity$            & 2.95E-48  & 0.56         \\
$n_{valid}$           & 8.20E-64  & 0.57         \\
$area_{in}$           & 8.05E-87  & 0.58          \\
$n_{in}$              & 5.06E-106 & 0.59         \\
$area_{cin}$          & 5.65E-119 & 0.60         \\
$n_{cin}$             & 3.97E-195 & 0.62       \\
$area_{out}$          & 0.00E+00  & 0.75        \\
$n_{out}$             & 0.00E+00  & 0.76         \\
$n_{cout}$            & 0.00E+00  & 0.77         \\
$area_{cout}$         & 0.00E+00  & 0.79        
\end{tabular}

%% file: table_pred_classic.tex
%Table: prediction performances for CLASSIC features (9) 

number of object\\
\begin{tabular}{ll|l}
\textbf{cells} & \textbf{voids} & \textbf{$\leftarrow$ predicted as} \\ \hline
1014            & 392            & \textbf{cells}                   \\
40            & 168            & \textbf{voids}                  
\end{tabular}
\\
object pixels\\
\begin{tabular}{ll|l}
\textbf{cells} & \textbf{voids} & \textbf{$\leftarrow$ predicted as} \\ \hline
3138180            & 579846            & \textbf{cells}                   \\
108800            & 448603            & \textbf{voids}                  
\end{tabular}

%% file: table_pred_all.tex
%Table: prediction performances for ALL features (19) 

number of object\\
\begin{tabular}{ll|l}
\textbf{cells} & \textbf{voids} & \textbf{$\leftarrow$ predicted as} \\ \hline
1127            & 279            & \textbf{cells}                   \\
20            & 188            & \textbf{voids}                  
\end{tabular}
\\
objects pixels\\
\begin{tabular}{ll|l}
\textbf{cells} & \textbf{voids} & \textbf{$\leftarrow$ predicted as} \\ \hline
3212541            & 505485            & \textbf{cells}                   \\
45702            & 511701            & \textbf{voids}                  
\end{tabular}

%% file: table_pred_ext.tex
%Table: prediction performances for EXT+2 features (11) 

number of object\\
\begin{tabular}{ll|l}
\textbf{cells} & \textbf{voids} & \textbf{$\leftarrow$ predicted as} \\ \hline
1068            & 338            & \textbf{cells}                   \\
27            & 181            & \textbf{voids}                  
\end{tabular}
\\
object pixels\\
\begin{tabular}{ll|l}
\textbf{cells} & \textbf{voids} & \textbf{$\leftarrow$ predicted as} \\ \hline
3142960            & 575066            & \textbf{cells}                   \\
61145            & 496258            & \textbf{voids}                  
\end{tabular}

%% file: table_p-value.tex
\begin{tabular}{|l|l|l|}
\hline
p-value         & significant   & all    \\ \hline
generic         & .01         & 3.8E-9 \\ \hline
significant     & -             & 3.8E-7  \\ \hline
\end{tabular}